\documentclass{iopart}
\usepackage{iopams,amssymb,psfig,graphicx}
\begin{document}
\title{Remote state preparation and teleportation in phase space}
\author{Matteo G. A. Paris} \address{Quantum Optics $\&$
Information Group, Unit\'a INFM di Pavia, Italia \\{\sf E-mail
address: paris@unipv.it}, {\sf URL: www.quantumoptics.it/\~{}paris}}
\author{Mary Cola and Rodolfo Bonifacio} 
\address{Dipartimento di Fisica and Unit\'a INFM, Universit\'a di Milano, Italia}
%%%%%%%%%%%%%%%%%%%%%%%%%%%%%%%%%%%%%%%%%%%%%%%%%%%%%%%%%%%%%%%%%%
\begin{abstract}
Continuous variable remote state preparation and teleportation are analyzed
using Wigner functions in phase space. We suggest a remote squeezed state 
preparation scheme between two parties sharing an entangled twin beam, where 
homodyne detection on one beam is used as a conditional source of squeezing 
for the other beam. The scheme works also with noisy measurements, and 
provide squeezing if the homodyne quantum efficiency is larger than $50\%$. 
Phase space approach is shown to provide a convenient framework to describe 
teleportation as a generalized conditional measurement, and to evaluate relevant 
degrading effects, such the finite amount of entanglement, the losses along
the line, and the nonunit quantum efficiency at the sender location.
\end{abstract}
%%%%%%%%%%%%%%%%%%%%%%%%%%%%%%%%%%%%%%%%%%%%%%%%%%%%%%%%%%%%%%%%%%
\section{Introduction}\label{s:intro}
Let us consider an entangled state described by a density 
matrix $R$ on a bipartite Hilbert space ${\cal H}_1 \otimes {\cal H}_2$.
A measurement performed on one subsystem reduces the other one according
to the projection postulate. Each possible outcome, say $x$, occurs with
probability $p_x$, and corresponds to a different conditional state
$\varrho_x$ 
\begin{eqnarray}
p_x = \hbox{Tr}_{12} \left[R\: \Pi_x \otimes I_2 \right]\:,  
\qquad \varrho_x = \frac1{p_x} \hbox{Tr}_1 \left[ R \: \Pi_x \otimes I_2\right]
\label{cond}\:.
\end{eqnarray}
$\Pi_x$ is the probability measure (POVM) of the measurement (acting on the
Hilbert space of the first subsystem) and $I_2$ the identity operator on the second
Hilbert space. $\hbox{Tr}_{12}\left[...\right]$ denotes full trace, whereas 
$\hbox{Tr}_j\left[...\right]$, $j=1,2$ denotes partial traces. \par
Eq. (\ref{cond}) shows that entanglement and conditional measurements can be
powerful resources to realize (probabilistically) nonlinear dynamics that
otherwise would not have been achievable through Hamiltonian evolution in
realistic media. Since entanglement may be shared between two distant
users (the sender performing the measurement, and the receiver observing the
conditional output), the inherent nonlocality of entangled 
states permits the {\em remote preparation} of the conditional states
$\varrho_x$, a protocol that may be used to exchange quantum 
information between the two parties sending only classical bits \cite{lo}.
A different kind of remote state preparation is teleportation \cite{tel},
where the measurement depends on an unknown reference state which may be
recovered at the receiver location {\em independently} on the outcome of the
measurement.  \par 
In this paper, we focus our attention on continuous variable (CV) remote
state preparation . In particular, we analyze in detail an optical scheme 
for remote preparation of squeezed states by realistic (noisy) conditional 
homodyining. Our analysis is based on a phase-space approach, and this is 
motivated by the following reasons: i) entanglement in optical CV quantum 
information processing is provided by the so-called twin-beam (TWB) state 
of two field modes $|\lambda\rangle\rangle=\sqrt{1-\lambda^2}\sum_p \lambda^p 
\: |p\rangle |p\rangle$, $0<\lambda<1$; the corresponding Wigner function
is Gaussian; ii) trace operation corresponds to overlap integral \cite{cah}, 
and the Wigner function of (realistic) homodyne POVM is also a Gaussian. 
By Wigner calculus we will be able to derive 
simple analytical formulas for conditional outputs, also in the case of 
noisy measurement at the sender location. In addition, we will 
show that phase-space approach is a convenient framework to describe 
CV teleportation as a conditional measurement , and to 
evaluate relevant degrading effects, such the finite amount of entanglement, 
the losses along the transmission channel, and the nonunit quantum efficiency 
at the sender location.
%%%%%%%%%%%%%%%%%%%%%%%%%%%%%%%%%%%%%%%%%%%%%%%%%%%%
\section{Conditional measurement in phase space}\label{s:cps}
TWB is the maximally entangled state (for a given, finite, value
of energy) of two modes of radiation. It can be produced either
by mixing two single-mode squeezed vacuum (with orthogonal
squeezing phases) in a balanced beam splitter \cite{kim} or, from the 
vacuum, by spontaneous downconversion in a nondegenerate parametric 
optical amplifier (NOPA) \cite{kum0}. The
evolution operator of the NOPA reads as follows $U_r =
\exp{\left[r \left(a^\dag b^\dag-ab\right)\right]}$ where the
"gain" $r$ is proportional to the interaction-time, the nonlinear
susceptibility, and the pump intensity.  We have $\lambda=\tanh
r$, whereas the number of photons of TWB is given by $N=2\sinh^2
r=2\lambda^2/(1-\lambda^2)$. In view of the duality
squeezing/entanglement via balanced beam-splitter \cite{joint} the parameter
$r$ is sometimes referred to as the squeezing parameter of the
twin-beam. Throughout the paper we will refer to mode $a$ as "mode 1" and
to mode $b$ as "mode 2".  The Wigner function 
$W[\hbox{\footnotesize TWB}](x_1,y_1;x_2,y_2)$ of a
TWB is Gaussian, and is given by (we omit the argument)
$$
W[\hbox{\footnotesize TWB}]=\left(2\pi \sigma_+^2 \: 
2\pi \sigma_-^2\right)^{-1}\:
\exp\left[-\frac{(x_1+x_2)^2}{4\sigma_+^2}
-\frac{(y_1+y_2)^2}{4\sigma_-^2} -\frac{(x_1-x_2)^2}{4\sigma_-^2}
-\frac{(y_1-y_2)^2}{4\sigma_+^2}\right]
$$
where the variances are given by  
\begin{eqnarray}
\sigma^2_+=\frac14\exp\{2r\} \qquad 
\sigma^2_-=\frac14\exp\{-2r\}
\label{stwb}\;.
\end{eqnarray}
Specializing Eq. (\ref{cond}) for $R=|\lambda\rangle\rangle\langle\langle\lambda |$ we have
\begin{eqnarray}
p_x &=& \langle\langle \lambda|\Pi_x\otimes I_2
|\lambda\rangle\rangle = (1-\lambda^2) \hbox{Tr}_1 \left[ 
\lambda^{a^\dag a} \: \Pi_x\right]\nonumber  \\
\varrho_x &=& \frac1{p_x} \hbox{Tr}_1 \left[|\lambda\rangle\rangle\langle\langle \lambda|\:
\Pi_x\otimes I_2\right]
\label{cond1}\;,
\end{eqnarray}
where, in the expression of $p_x$, we have already 
performed the trace over the Hilbert space ${\cal H}_2$. 
In the following, the partial traces in Eq. (\ref{cond1}) will be
evaluated as overlap integrals in the phase space. The Wigner
function of a generic operator $O$ is defined as the following
complex Fourier transform
\begin{eqnarray}
W[O](\alpha) &=& \int \frac{d^2\gamma}{\pi^2}
e^{\alpha\bar\gamma - \bar\alpha\gamma}\: \hbox{Tr}\left[O\: D(\gamma)\right]
\label{Wigs}\;,
\end{eqnarray}
where $\alpha$ is a complex number, and $D(\gamma)=e^{\gamma a^\dag - \bar\gamma a}$ is the 
displacement operator. 
The inverse transformation reads as follows \cite{msacchi}
\begin{eqnarray}
O =  \int d^2\alpha \: W[O] (\alpha)\:  e^{-2 |\alpha|^2} \: e^{2\alpha a^\dag}
\left(-\right)^{a^\dag a} e^{2\bar\alpha a}
\label{invwig}\;
\end{eqnarray}
Using the Wigner function the trace
between two operators can be written as 
\begin{eqnarray}
\hbox{Tr}\left[O_1\: O_2\right] &=& \pi \int d^2\beta\:W[O_1](\beta)\:W[O_2](\beta)
\label{trace}\;.
\end{eqnarray}
%%%%%%%%%%%%%%%%%%%%%%%%%%%%%%%%%%%%%%%%%%%%%%%%%%%%
\subsection{Remote squeezed states preparation}
\begin{figure}[h]
\psfig{file=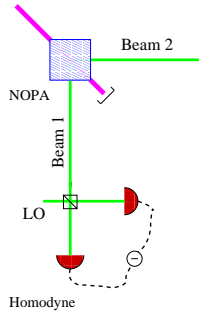,width=30mm}
\caption{Schematic diagram of conditional homodyne for remote squeezed state
preparation. A TWB is produced by spontaneous downconversion in a NOPA, and
then homodyne detection is performed on one of the two modes, say
mode $1$. Mode $2$ is squeezed if the homodyne quantum efficiency is larger
than $50\%$.\label{f:rsp}}
\end{figure}

Let us consider the optical scheme depicted in Fig. \ref{f:rsp}.
A TWB is produced by spontaneous downconversion in a NOPA, and
then homodyne detection is performed on one of the two modes, say
mode $1$. The POVM of the measurement, assuming perfect
detection {\em i.e.} unit quantum efficiency, is given by 
\begin{eqnarray}
\Pi_x = |x\rangle\langle x| \qquad\qquad |x\rangle = \left(\frac2\pi\right)^{1/4}
e^{-2x^2} \sum_p \frac{H_p(\sqrt{2}x)}{\sqrt{2^p p!}} |p\rangle
\label{homod}\;,
\end{eqnarray}
$|x\rangle$'s being eigenstates of the quadrature operator $x=1/2
(a + a^\dag)$. The Wigner function of the POVM
$\Pi_x$ is a delta function 
\begin{eqnarray}
W[\Pi_x] (x_1) = \delta (x_1-x) 
\label{pomx}\;,
\end{eqnarray}
whereas that of the term $\lambda^{a^\dag a}$ in the first of Eqs. 
(\ref{cond1}) is  given by 
\begin{eqnarray}
(1-\lambda^2) W[\lambda^{a^\dag a}] (x_1,y_1) &=& (2\pi\sigma^2)^{-1}
\exp\left\{-\frac{x_1^2+y_1^2}{2\sigma^2}\right\} \label{lam}\:,
\end{eqnarray}
where the variance $\sigma$ depends on the number of photons of the TWB 
$\sigma^2=\frac14 (1+N)$. Using Eqs. (\ref{pomx}) and (\ref{lam}) 
it is straightforward to evaluate the probability distribution 
\begin{eqnarray}
p_x &=& \!\!\int\!\!\!\!\int\!\!\!\!\int\!\!\!\!\int\!\!  
dx_1 dy_1 dx_2 dy_2 \: W[\hbox{\footnotesize TWB}](x_1,y_1;x_2,y_2) 
W[\Pi_x] (x_1) \nonumber \\ &=&
(1-\lambda^2) \!\!\int\!\!\int\!\!\ dx_1 dy_1 \: 
W[\lambda^{a^\dag a}](x_1,y_1)\:W[\Pi_x] (x_1) 
\nonumber \\ &=&
(2\pi\sigma^2)^{-1/2} \exp\left\{-\frac{x^2}{2\sigma^2}\right\} 
\label{px}\;,
\end{eqnarray}
and the Wigner function of the conditional output state 
\begin{eqnarray}
W[\varrho_x] (x_2,y_2) &=& 
\!\!\int\!\!\!\!\int\!\! dx_1 dy_1 \: W[\hbox{\footnotesize TWB}](x_1,y_1;x_2,y_2) 
W[\Pi_x] (x_1) \nonumber \\ &=& (2\pi\Sigma_1^2 \: 2\pi\Sigma_2^2)^{-1/2} 
\exp\left\{-\frac{(x_2-a_x)^2}{2\Sigma_1^2}
-\frac{y_2^2}{2\Sigma_2^2} \right\} 
\label{condx}\;.
\end{eqnarray}
The parameters in Eq. (\ref{condx}) are given by
\begin{eqnarray}
a_x = \frac{\sqrt{N(N+2)}}{1+N} x\qquad
\Sigma_1^2 = \frac14 \frac1{1+N} \quad
\Sigma_2^2 = \frac14 (1+N)
\label{parsx}\;.
\end{eqnarray}
Eqs. (\ref{condx}) and (\ref{parsx}) say that $\varrho_x$ is a squeezed-coherent 
minimum uncertainty state of the form $\varrho_x = D(a_x) S(r_x) |0\rangle$ 
{\em i.e.} a state squeezed in the direction of the measured quadrature 
$\overline{\Delta x^2}=1/4 e^{-2 r_x}$, with
squeezing parameter given by $r_x = 1/2 \log (1+N)$. Notice that this result 
is valid for any quadrature $x_\phi=e^{ia^\dag a \phi} x e^{-i a^\dag a \phi}$, 
and therefore the present scheme, by tuning the phase of the local oscillator 
in the homodyne detection, is suitable for the remote preparation of squeezed states
with any desired phase of squeezing. Of course, we have squeezing for $\varrho_x$
if and only if  $N>0$ {\em i.e.} if and only if  entanglement is present. \par
A question arises whether or not the remote preparation of squeezing is
possible with realistic homodyne detection, {\em i.e.} with noisy measurement
of the field quadrature. The POVM of a homodyne detector with quantum efficiency
$\eta$ is a Gaussian convolution of the ideal POVM 
\begin{eqnarray}
\Pi_{x\eta} = \int \frac{dy}{\sqrt{2\pi\sigma_\eta^2}}
\: \exp\{-\frac{(y-x)^2}{2\sigma_\eta^2}\} \: |y\rangle\langle y| 
\label{homodeta}\;,
\end{eqnarray}
with $\sigma_\eta^2=\frac14 (1-\eta)/\eta$ \cite{paul}. 
The corresponding Wigner function is given by
\begin{eqnarray}
W[\Pi_{x\eta}] (x_1) = (2\pi\sigma_\eta^2)^{-1/2} 
\exp\left\{-\frac{(x_1-x)^2}{2\sigma_\eta^2}
\right\}
\label{wighometa}\;.
\end{eqnarray}
Using (\ref{wighometa}) one evaluates the probability distribution and 
the Wigner function of the conditional output state, one has 
\begin{eqnarray}
p_{x\eta} = [2\pi(\sigma^2+\sigma_\eta^2)]^{-1/2}
\exp\left\{-\frac{x^2}{2(\sigma^2+\sigma_\eta^2)}\right\} \\
\fl W[\varrho_{x\eta}] (x_2,y_2) = 
(2\pi\Sigma_{1\eta}^2 \: 2\pi\Sigma_{2\eta}^2)^{-1/2} 
\exp\left\{-\frac{(x_2-a_{x\eta})^2}{2\Sigma_{1\eta}^2}
-\frac{y_2^2}{2\Sigma_{2\eta}^2}
\right\}\label{condxeta0}\;,
\end{eqnarray}
where
\begin{eqnarray}
\fl a_{x\eta} = \frac{\eta\sqrt{N(N+2)}}{1+\eta N} x\qquad
\Sigma_{1\eta}^2 = \frac14 \frac{1+N(1-\eta)}{1+\eta N} \quad
\Sigma_{2\eta}^2 = \frac14 (1+N)
\label{condxeta1}\;.
\end{eqnarray}
As a matter of fact, the conditional output $\varrho_{x\eta}$ is no longer 
a minimum uncertainty state. However, for $\eta$ large enough, it still shows 
squeezing in the direction individuated by the measured quadrature {\em i.e.} 
$\overline{\Delta x^2}<1/4$. In order to obtain the explicit form of the conditional 
output state from the Wigner function $W[\varrho_{x\eta}] (x_2,y_2)$ of Eq.
(\ref{condxeta0}) we use Eq. (\ref{invwig}) arriving at 
\begin{eqnarray}
\varrho_{x\eta}= D(a_{x\eta})S(r_{x\eta}) \nu_{th} S^\dag(r_{x\eta}) D^\dag (a_{x\eta})
\label{outxeta}\;,
\end{eqnarray}
where $\nu_{th}= (1+n_{th})^{-1} \sum_p [n_{th}/(1+n_{th})]^p \:
|p\rangle\langle p|$ is a thermal state with average number of photons given
by 
\begin{eqnarray}  
n_{th} &=& \frac12 \left\{\sqrt{\frac{(1+N)[1+N(1-\eta)]}{1+\eta N}}-1\right\}
\label{ntheta}\;,
\end{eqnarray}
and the squeezing parameter is given by
\begin{eqnarray}  
\\  r_{x\eta} &=& \frac14 \log \frac{(1+N)(1+\eta N)}{1+N(1-\eta)}
\label{rxeta}\;.
\end{eqnarray}
We have squeezing in $\varrho_{x\eta}$ if $\Sigma_{1\eta}^2<1/4$, and this happens
for $\eta > 50 \%$ independently on the actual value $x$ of the homodyne outcome. 
The values of efficiency that can be currently realized in a quantum
optical lab is far above the $50\%$ limit, and thus we conclude that 
conditional homodyning on TWB is a robust scheme for the remote preparation of squeezing.
%%%%%%%%%%%%%%%%%%%%%%%%%%%%%%%%%%%%%%%%%%%%%%%%%%%%
\subsection{Teleportation as a generalized conditional measurement}
\begin{figure}[h]
\psfig{file=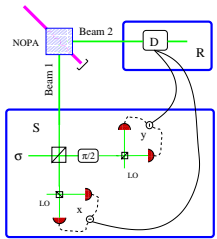,width=30mm}
\caption{Schematic diagram of teleportation scheme.  In the sender area (S), one
part of a TWB is mixed with a given reference state $\sigma$ (the state to be
teleported) in a balanced beam splitter, and two orthogonal quadratures are
measured on the outgoing beams by means of two homodyne detectors with local
oscillators phase-shifted by $\pi/2$.  After the measurement, in the receiver
area (R), the other part of the TWB is displaced by an amount $-\alpha=-x-iy$ 
that depends on the outcome of the measurements itself. The  overall state,
averaged over the possible outcomes, is the teleported state. 
\label{f:tel}}
\end{figure}

The scheme for optical CV teleportation is depicted in Fig. \ref{f:tel}.
One part of a TWB is mixed with a given reference state $\sigma$ in a balanced
beam splitter, and two orthogonal quadratures $x=1/2 (a+a^\dag)$, 
$y=i/2 (a^\dag -a)$ are measured on the outgoing beams by means of two
homodyne detectors with local oscillators phase-shifted by $\pi/2$. 
The other part of the TWB is then displaced by an amount $-\alpha=-x-iy$ that 
depends on the outcome of the measurements, and the resulting state 
(averaged over the possible outcomes) is the teleported state. \par 
Overall, the measurement performed on the TWB 
is a generalized double homodyne detection \cite{tritter,busch} (equivalent to generalized
heterodyne), which can be described by the POVM \cite{busch,iccsur}
\begin{eqnarray}
\Pi_\alpha = D(\alpha) \sigma^T D^\dag (\alpha)
\label{pialpha}\;,
\end{eqnarray}
$...^T$ denoting transposition. Therefore, using Eq. (\ref{cond1}), 
one has 
\begin{eqnarray}
p_\alpha &=& \langle\langle \lambda|\Pi_\alpha\otimes I_2
|\lambda\rangle\rangle = (1-\lambda^2) \hbox{Tr}_1 \left[ 
\lambda^{a^\dag a} \: D(\alpha) \sigma^T D^\dag (\alpha)
\right]\nonumber  \\
\varrho_\alpha &=& \frac1{p_\alpha} D(-\alpha)\hbox{Tr}_1 
\left[|\lambda\rangle\rangle\langle\langle \lambda|\:
D(\alpha) \sigma^T D^\dag (\alpha)
\otimes I_2\right] D^\dag (-\alpha)
\label{condalpha}\;,
\end{eqnarray}
while the teleported state is given by
\begin{eqnarray}
\fl \varrho = \int d^2\alpha \: p_\alpha \: \varrho_\alpha
= \int d^2\alpha \:
D(-\alpha)\hbox{Tr}_1 
\left[|\lambda\rangle\rangle\langle\langle \lambda|\:
D(\alpha) \sigma^T D^\dag (\alpha)
\otimes I_2\right] D^\dag (-\alpha)
\label{teleported}\;.
\end{eqnarray}
Using Wigner functions and taking into account that for any density matrix
\begin{eqnarray}
W[\varrho^T](x,y) = W[\varrho](x,-y) \\  
W[D(\alpha)\varrho D^\dag (\alpha)](x,y) = W[\varrho](x-x_\alpha,y-y_\alpha) 
\label{rems}\;,
\end{eqnarray}
with $x_\alpha=\hbox{Re} [\alpha]$ and $y_\alpha=\hbox{Im}[\alpha]$, one has 
\begin{eqnarray}
\fl W[\varrho](x_2,y_2) = \nonumber \\ 
\fl=\!\!\int\!\!\!\!\int\!\!  dx_1 dy_1 \!\!
 \int\!\!\!\!\int\!\! dx_\alpha dy_\alpha
\:  W[\hbox{\footnotesize
TWB}](x_1,y_1;x_2+x_\alpha,y_2+y_\alpha) \: W[\sigma] 
(x_1-x_\alpha,-y_1-y_\alpha) \nonumber \\
\fl=\!\!\int\!\!\!\!\int\!\!  dx_1 dy_1\: W[\sigma] (x_1,y_1) 
 \int\!\!\!\!\int\!\! dx_\alpha dy_\alpha
\:  W[\hbox{\footnotesize
TWB}](x_1+x_\alpha,-y_1-y_\alpha;x_2+x_\alpha,y_2+y_\alpha) 
\nonumber \\
\fl= \!\!\int\!\!\!\!\int\!\!  \frac{dx_1 dy_1}{\pi \kappa_r^2}
\exp\left\{-\frac{(x_1-x_2)^2+(y_1-y_2)^2}{\kappa^2_r}\right\}\: 
W[\sigma] (x_1,y_1)
\nonumber \\
\fl= \!\!\int\!\!\!\!\int\!\!  \frac{dx_1 dy_1}{\pi \kappa_r^2}
\exp\left\{-\frac{x_1^2+y_1^2}{\kappa^2_r}\right\}\: 
W[D(\alpha_1)\sigma D^\dag (\alpha_1)] (x_2,y_2)\label{wigtel}\;,
\end{eqnarray}
with $\alpha_1=x_1+ iy_1$ and $\kappa_r^2=\exp\{-2 r\}$. 
From Eqs. (\ref{wigtel}) and 
(\ref{invwig})  one has that the teleported state is given by 
\begin{eqnarray}
\varrho = \int \frac{d^2\alpha}{\pi\kappa_r^2}\: 
\exp\{-\frac{|\alpha|^2}{\kappa_r^2}\}
\: D(\alpha)\sigma D^\dag (\alpha)
\label{tel}\;,
\end{eqnarray}
which coincides with the input state only in the limit $r\longrightarrow
\infty$ {\em i.e.} for infinite energy of the TWB. Eq. (\ref{tel}) shows that 
CV teleportation with finite amount of entanglement is equivalent to a
thermalizing channel with $\kappa_r$ thermal photons: this results 
has been obtained also with other methods \cite{ban}. However, the present 
Wigner approach may be more convenient in order to include other degrading 
effects such the nonunit quantum efficiency at the sender location 
and the losses along the transmission channel. \par
Nonunit quantum efficiency at the homodyne detectors affects the POVM of the 
sender, which become a Gaussian convolution of the ideal POVM $\Pi_\alpha$
\begin{eqnarray}
\Pi_{\alpha\eta} = \int \frac{d^2\beta}{\pi\Delta_\eta^2}
\: \exp\{-\frac{|\alpha - \beta|^2}{\Delta_\eta^2}\} \: \Pi_\beta 
\label{pialfaeta}\;,
\end{eqnarray}
with $\Delta_\eta^2=(1-\eta)/\eta$ \cite{paul}. 
On the other hand, losses along the line degrade the
entanglement of the TWB supporting the teleportation. The propagation of a TWB
inside optical media can be modeled as the coupling of each part of the 
TWB with a non zero temperature reservoir. 
The dynamics can be described in terms of the two-mode Master equation
\fl \begin{eqnarray} \fl
\frac{d\varrho_t}{dt} \equiv {\cal L} \varrho_t = 
\Gamma (1+M) L[a] \varrho_t + \Gamma (1+M) L[b] \varrho_t  + 
\Gamma M L[a^\dag ] \varrho_t + \Gamma  M  L[b^\dag] \varrho_t  
\label{master} \end{eqnarray}
where $\varrho_t\equiv\varrho (t)$, $\Gamma$ 
denotes the (equal) damping rate, $M$ the number of 
background thermal photons, and $L[O]$ is the Lindblad superoperator 
$L[ O ] \varrho_t =  O \varrho_t  O^\dag - \frac{1}{2} O^\dag  O 
\varrho_t - \frac{1}{2} \varrho_t OO^\dag\:.$
The terms proportional to $L[a]$ and  $L[b]$ describe the losses, 
whereas the terms proportional to $L[a^\dag]$ and $L[b^\dag]$ describe 
a linear phase-insensitive amplification process. 
This can be due either to optical media dynamics or to thermal 
hopping; in both cases no phase information is carried.
Of course, the dissipative dynamics of the two channels are independent on each other. 
The master equation (\ref{master}) can be transformed into a Fokker-Planck
equation for the two-mode Wigner function of the TWB
Using the differential representation of the superoperators in Eq. 
(\ref{master}) the corresponding Fokker-Planck equation reads as follows 
\begin{eqnarray}
\fl\partial_\tau W_\tau = \left[ \frac{1}{8}%
\left(\sum_{j=1}^2\partial^2_{x_j x_j} + \partial^2_{y_j y_j}\right) 
+ \frac{\gamma}2 \left(\sum_{j=1}^2 \partial_{x_j} x_j+\partial_{y_j}  y_j  
\right) \right] W_\tau \label{fp}\:,  
\end{eqnarray}
where $\tau$ denotes the rescaled time $\tau=(\Gamma/\gamma)\:t$, and 
$\gamma= \frac{1}{2M+1}$ the drift term.
The solution of Eq. (\ref{fp}) can be written as 
\begin{eqnarray}\fl
W_\tau &=& \int _{}dx^{\prime}_1\int
_{}dx^{\prime}_2 \int _{}dy^{\prime}_1\int _{}dy^{\prime}_2 \;\:
W[\hbox{\footnotesize TWB}](x^{\prime}_1,y^{\prime}_1;x^{\prime}_2,y^{\prime}_2)\: 
\prod_{j=1}^2 G_\tau(x_j|x^{\prime}_j) 
G_\tau(y_j|y^{\prime}_j) \: 
\label{conv}
\end{eqnarray}
where $W[\hbox{\footnotesize TWB}]$ is initial Wigner function of the 
TWB, and the Green functions $G_\tau(x_j|x^{\prime}_j)$ are given by 
\begin{eqnarray}\fl
G_\tau(x_j|x^{\prime}_j)=\frac{1}{\sqrt{2\pi D^2}}\exp\left[-\frac{
(x_j-x^{\prime}_je^{-\frac12 \gamma \tau})^2} {2 D^2}\right]
\;,\quad D^2=\frac{1}{4\gamma }(1-e^{-\gamma \tau}) \;.
\label{green}
\end{eqnarray}
The Wigner function $W_\tau$ can be obtained 
by the convolution (\ref{conv}), which can be easily evaluated since the initial 
Wigner function is Gaussian. The form of $W_\tau$ is the same of 
$W[\hbox{\footnotesize TWB}]$ with the variances changed to  
\begin{eqnarray}
\sigma_+^2 \longrightarrow \left(e^{-\gamma\tau}\sigma_+^2+D^2\right)\qquad 
\sigma_-^2 \longrightarrow \left(e^{-\gamma\tau}\sigma_-^2+D^2\right)
   \label{evolvedVar}\:.
\end{eqnarray}
Inserting the Wigner functions of the blurred POVM $\Pi_{\alpha\eta}$
and of the evolved TWB in Eqs. (\ref{condalpha}) and (\ref{teleported}) 
we obtain the teleported state in the general case, 
which is still given by Eq. (\ref{tel}), with the parameter $\kappa_r$
now given by 
\begin{eqnarray}
\kappa_r^2 \longrightarrow e^{-\Gamma t - 2 r} + (2M+1)(1-e^{-\Gamma t}) +
\Delta_\eta^2
\label{kappar}\;.
\end{eqnarray}
Eqs. (\ref{tel}) and (\ref{kappar}) summarize all the degrading effects
on the quality of the teleported state. In the special case of coherent 
state teleportation $\sigma=|z\rangle\langle z|$ (which corresponds to original 
optical CV teleportation experiments \cite{kim})
the fidelity $F=\langle z|\varrho |z\rangle$ can be evaluated straightforwardly 
as the overlap of the Wigner functions. Since 
$W[z](\alpha)=2/\pi\: e^{-2|\alpha -z|^2}$ is 
the Wigner function of a coherent state we have 
$$ F = \frac{1}{1+e^{-2r-\Gamma t}+(1-e^{-\Gamma t})(2M+1)+(1-\eta)/\eta}\:.$$ 
The condition on the fidelity, in order to assure that the scheme is a truly
nonlocal protocol, is given by $F>1/2$ \cite{kim}, {\em i.e.} 
$$ e^{-2r-\Gamma t}+(1-e^{-\Gamma t})(2M+1)+(1-\eta)/\eta <1\:. $$
Therefore, the bound on the quantum efficiency to demonstrate quantum
teleportation is given by 
$$ \eta > \frac{1}{2-e^{-2r-\Gamma t}-(1-e^{-\Gamma t})(2M+1)}\:.$$
If the propagation induces low  perturbation {\em i.e.} 
if $\Gamma\simeq 0$ and $M\simeq0$ we have $\eta > (2-e^{-2r})^{-1}$, 
which ranges from $1/2$ to $1$, and represents the range of "useful" 
values for the quantum efficiency. If $\Gamma$ and $M$ are not negligible 
then, for the same initial squeezing, we need a larger value of the 
quantum efficiency. Moreover, since quantum efficiency should be 
lower or equal to unit $\eta\leq1$, we may derive a bound 
on the initial squeezing that allows to demonstrate quantum 
teleportation. This reads as follows
$e^{-2r}\leq(2M+1)-2Me^{\Gamma t}$. Remarkably, if the number of thermal photons 
is zero, {\em i.e.} if the TWB is propagating in a zero temperature
environment, then any value of the initial squeezing parameter make
teleportation possible, of course if the quantum efficiency at the 
receiver location satisfies $\eta\geq(2 - e^{-2r -\Gamma t}-1+e^{\Gamma
t})^{-1}$.
%%%%%%%%%%%%%%%%%%%%%%%%%%%%%%%%%%%%%%%%%%%%%%%%%%%%
\section{Conclusions}\label{s:outro}
A method for the remote preparation of squeezed states by conditional
homodyninig on a TWB has been suggested. The scheme has been studied using
Wigner function, which is the most convenient approach to describe effects of
nonunit quantum efficiency at homodyne detectors.  The method is shown to
provide remote squeezing if the quantum efficiency is larger than $50\%$.
Since downconversion correlates pair of modes at any frequencies $\omega_1$
and $\omega_2$ satisfying $\omega_1+\omega_2=\omega_P$, $\omega_P$ being the
frequency of the pump beam, the present method can be used to generate
squeezing at frequencies where no media for degenerate downconversion are
available \cite{tun}. \par 
Phase-space approach has been also used to analyze
CV teleportation as a conditional generalized double homodyning on a TWB. Also
in this case the use of Wigner functions represents a powerful tool to
evaluate the degrading effects of finite amount of entanglement, losses along
the transmission channel, and nonunit quantum efficiency at sender location.
A bound on the value of quantum efficiency needed to demonstrate quantum
teleportation has been derived.
%%%%%%%%%%%%%%%%%%%%%%%%%%%%%%%%%%%%%%%%%%%%%%%%%%%%
\section*{Acknowledgments} This work has been sponsored by the INFM through 
the project PRA-2002-CLON, by MIUR through the PRIN projects {\em Decoherence
control in quantum information processing} and {\em Entanglement assisted high
precision measurements}, and by EEC through the project IST-2000-29681 (ATESIT).
MGAP is research fellow at {\em Collegio Alessandro Volta}.
%%%%%%%%%%%%%%%%%%%%%%%%%%%%%%%%%%%%%%%%%%%%%%%%%%%%%%%%%%%%%%%%%

%%%%%%%%%%%%%%%%%%%%%%%%%%%%%%%%%%%%%%%%%%%%%%%%%%%%%%%%%%%%%%%%%%
%%%%%%%%%%%%%%%%%%%%%%%%%%%%%%%%%%%%%%%%%%%%%%%%%%%%%%%%%%%%%%%%%%
\end{document}